\documentclass[twocolumn,aps,prb,superscriptaddress,a4paper,nofootinbib,showpacs,showkeys,lengthcheck]{revtex4-1}

\setcitestyle{numbers,square}

\usepackage{graphicx}
\usepackage{type1cm}
\usepackage[normalem]{ulem}  

\usepackage{bm,dcolumn,amssymb,amsmath,braket,siunitx,color}
\usepackage{sidecap}

\newcommand{\mm}[1]{\mbox{$#1$}}

\newcommand{\sigxx}{\mbox{$\sigma_{xx}$}}
\newcommand{\sigxy}{\mbox{$\sigma_{xy}$}}
\newcommand{\sigZxy}{\mbox{$\sigma^{z}_{xy}$}}

\newcommand{\rhoxx}{\mbox{$\rho_{xx}$}}
\newcommand{\rhoxy}{\mbox{$\rho_{xy}$}}

\newcommand{\rhoav}{\mbox{$\rho_{\text{aver}}$}}

\newcommand{\ef}{\mbox{$E_{F}$}}

\newcommand{\ea}{{\it et al.}}

\DeclareMathOperator{\TR}{Tr}

\begin{document}

\title{Transport properties of doped permalloy via ab-initio
  calculations: effect of the host disorder}



\author{O. \surname{\v{S}ipr}} 
\email{sipr@fzu.cz}
\homepage{http://www.fzu.cz/~sipr} \affiliation{ Institute of Physics
  of the Czech Academy of Sciences, Cukrovarnick\'{a}~10,
  CZ-162~53~Prague, Czech Republic }  \affiliation{New Technologies Research
  Centre, University of West Bohemia, CZ-301~00~Pilsen, Czech Republic}

\author{S. \surname{Wimmer}} \affiliation{Universit\"{a}t
  M\"{u}nchen, Department Chemie, Butenandtstr.~5-13,
  D-81377~M\"{u}nchen, Germany}

\author{S. \surname{Mankovsky}} \affiliation{Universit\"{a}t
  M\"{u}nchen, Department Chemie, Butenandtstr.~5-13,
  D-81377~M\"{u}nchen, Germany}

\author{H. \surname{Ebert}} \affiliation{Universit\"{a}t M\"{u}nchen,
  Department Chemie, Butenandtstr.~5-13, D-81377~M\"{u}nchen, Germany}

\date{\today}

\begin{abstract}
Transport properties of permalloy doped with V, Co, Pt, and Au are
explored via ab-initio calculations.   For this purpose,  the
Kubo-Bastin formula is evaluated within the fully relativistic
Korringa-Kohn-Rostoker Green function formalism.  Finite temperature
effects are treated by means of the alloy analogy model.  It is shown
that the fact that the host is disordered and not crystalline has a
profound effect on how the conductivities characterizing the anomalous
Hall effect and the spin Hall effect depend on the dopant
concentration.  Several relationships between quantities
characterizing charge and spin transport are highlighted.  The rate of
decrease of the longitudinal charge conductivity with increasing
doping depends on the dopant type, following the sequence
Co--Au--Pt--V.  The dependence of the anomalous Hall and spin Hall
conductivities on the dopant concentration is found to be
non-monotonic.  Introducing a finite temperature changes the overall
trends significantly.  The theoretical results are compared with
available experimental data.
\end{abstract}



\maketitle


\section{Introduction}   \label{sec-intro}

There have been growing efforts to complement conventional
electronics, based on manipulating conducting electrons via charge, by
spintronics, which manipulates the electron spin.  A prominent role is
played in this respect by transport phenomena closely linked to the
spin orbit coupling (SOC), such as the anomalous Hall effect (AHE),
the spin Hall effect (SHE), or the anisotropic magnetoresistance
(AMR).  Intimately connected with the development of spintronics is
the search for corresponding new materials.  The modification of
magnetic materials by doping is a promising and intensively studied
way to make progress.  Transport properties can be strongly influenced
even by very low dopant concentrations.  Understanding the underlying
mechanisms responsible for the modification of transport properties is
thus very important.  A reliable ab-initio description of the way
doping affects transport phenomena is a necessary part of this
process.

So far the theoretical description in the field has focused mostly on
how doping influences transport properties of crystalline, i.e., clean
systems.  Concepts and intuitive views have been established that can
be conveniently used when interpreting experiments
\cite{CB01,OSN06,Sin08}.  However, experimental and technological
interest is turning also to alloys which offer a diverse range of
properties.  The question is to what extent the approaches
that proved to be useful for doped crystals can be transferred to
doped alloys.  To start with, one of the key factors affecting
theoretical analysis of transport in doped crystals is that the
longitudinal charge conductivity tends to infinity for zero doping
(for $T$=0~K).  This is no longer true if the host is a disordered
alloy.  So one can presume that some trends of the transport
properties with doping which are common and well understood for
crystalline hosts will not occur for doped alloys.

One of the widely studied chemically disordered materials is permalloy
Fe$_{19}$Ni$_{81}$.  It is attractive because of its high magnetic
permeability but also because of its transport properties, which are
characterized by a high and low electrical conductivity in the
majority and minority spin channels, respectively.  Several studies
how permalloy (Py) properties can be modified via doping by magnetic
or non-magnetic atoms were published recently
\cite{NST+00,LCK+10,LCC+10,YPA+15,PCH+16,HGS+16,DVS+18,SME+19}.

The aim of this study is to investigate transport properties of doped
Py via ab-initio calculations, for zero and finite temperatures alike.
Special attention is paid to understanding the trends with the doping
and to how this compares with trends for crystalline hosts.  Apart
from the longitudinal conductivity, we focus on the SOC-related
phenomena AHE, SHE, and AMR.  The dopants we consider are V, Co, Pt,
and Au.  In this way we cover a wide range of circumstances: E.g., Co
is magnetic, V and Pt are non-magnetic but easily polarizable, Au is
non-magnetic and hard to polarize.  The induced magnetic moment of V
and Pt is oriented antiparallel and parallel, respectively, to the
magnetization of the Py host.  The SOC is weak at V and Co but strong
at Pt and Au.  Co lies between Fe and Ni in the periodic table, so it
will presumably not disturb the electronic structure of Py much while
the other dopants should pose a substantial disturbance.

By performing ab-initio calculations and analyzing our results, we
will show that the fact that the Py host is disordered and not
crystalline has profound influence on the dependence of the AHE and
SHE on the dopant concentration.  In particular, this dependence
cannot be ascribed unambigously to skew scattering, side-jump
scattering, or intrinsic contributions in the same way as as it can be
done when investigating the effect of doping for an ordered
crystalline host.   Where possible, our theoretical results
are compared with available 
experimental data.



\section{Theoretical scheme}    \label{sec-method}


\subsection{Charge and spin conductivities}    \label{sec-kubo}

The charge and spin conductivities are calculated in a
linear-response regime, using a particular form of the Kubo-Bastin
equation~\cite{BLBN71}
implemented using the fully relativistic
multiple-scattering or Korringa-Kohn-Rostoker Green function (KKR-GF)
method \cite{EKM11}.  Introducing a generalized conductivity
$\mathcal{C}_{\mu\nu}$, we can express it as \cite{KCE15}
\begin{widetext}
\begin{eqnarray}
\mathcal{C}_{\mu\nu}
&=&
\label{eq-Bastin}
\mathcal{C}^{I}_{\mu\nu}
+\mathcal{C}^{II}_{\mu\nu}   \;\; ,
\\
\mathcal{C}^{I}_{\mu\nu}
&=&
\label{eq-Bastin-I}
\frac{\hbar }{4\pi\Omega}
\TR
\left<
\hat{O}_{\mu}(\hat{G}^{+}-\hat{G}^{-})  \hat{j}_{\nu} \hat{G}^{-}
-\hat{O}_{\mu} \hat{G}^{+} \hat{j}_{\nu}(\hat{G}^{+}-\hat{G}^{-})
\right>_{\! c}   \;\; ,
\\
\mathcal{C}^{II}_{\mu\nu}
&=&
\label{eq-Bastin-II}
\frac{\hbar }{4\pi\Omega}\int_{-\infty}^{ E_{\mbox{\tiny  F}}}
\TR
\left<
\hat{O}_{\mu} \hat{G}^{+}\hat{j}_{\nu}
\frac{d\hat{G}^{+}}{d E}
-\hat{O}_{\mu}\frac{d\hat{G}^{+}}{d E}  \hat{j}_{\nu} \hat{G}^{+}
-
\left(
\hat{O}_{\mu} \hat{G}^{-}            j_{\nu}\frac{d\hat{G}^{-}}{d E}
-\hat{O}_{\mu}\frac{d\hat{G}^{-}}{d E}  \hat{j}_{\nu} \hat{G}^{-}
\right)
\right>_{\! c}
{\rm d}
 E
\;\; ,
\end{eqnarray}
\end{widetext}
where $\hat{G}^{+}$ and $\hat{G}^{-}$ are the retarded and advanced 
Green functions, respectively, $\Omega$ is the unit cell volume, and
$\mu$, $\nu$ are the Cartesian coordinates.  If the generalized
conductivity $\mathcal{C}_{\mu\nu}$ stands for the the charge
conductivity $\sigma_{\mu\nu}$, the generalized current operator
$\hat{O}_{\mu}$ stands for the electric current operator
$\hat{j}_{\mu}$, given within the relativistic formalism as
\cite{Ros61} 
\begin{equation}
 \hat{\bm{j}} = -|e|c\mbox{\bm{$\alpha$}}
\; .
\end{equation}
If $\mathcal{C}_{\mu\nu}$ stands for the the spin Hall conductivity
$\sigma_{\mu\nu}^{z}$ (with the spin quantization axis along the $z$
coordinate), then $\hat{O}_{\mu}$ stands for the relativistic spin
current density operator $\hat{J}^{z}_{\mu}$, given by \cite{KCE15}
\begin{equation}
\hat{J}^{z}_{\mu} = \left(
  \beta \Sigma_{z} - \frac{\gamma_{5}\hat{p}_{z}}{mc}
  \right)
  |e| c \alpha_{\mu}
\; ,
\end{equation}
 where we restrict to the cases $\mu=x,y$.
The quantities $e$, $m$, and $c$ have their usual meaning,
$\hat{p}_{z}$ is the canonical momentum, $\bm{\alpha}$ and $\beta$ are
the standard 4$\times$4 matrices occurring in the Dirac formalism, and
$\Sigma_{z}$ and $\gamma_{5}$ are 4$\times$4 matrices defines as
\[
\Sigma_{z} = \left(
\begin{array}{cc}
  \sigma_{z}  &  0 \\
  0          &  \sigma_{z}
\end{array}
\right)
\; ,
\qquad
\gamma_{5} = \left(
\begin{array}{cc}
    0    &  -I_{2} \\
  -I_{2}  &  0
\end{array}
\right)
\; .
\]
For more details see, e.g., Refs.~\onlinecite{LKE10,LGK+11,KCE15}.

The angular brackets \mm{\left< \right>_{\! c}}\ in
Eqs.~(\ref{eq-Bastin-I}) and (\ref{eq-Bastin-II}) indicate averaging
over configurations, as is required when investigating alloys.  To
perform this averaging we employ the coherent potential approximation
(CPA).  This encompasses the self-consistent determination of the
single-site potentials as well as configurational averaging when
calculating transport properties.  In the latter case one has to
include the so-called vertex corrections~\cite{Vel69,But85}, which account
for the difference between a configurational average of a product and
a product of configurational averages,
\begin{equation}
\left<
\hat{O}_{\mu} \hat{G}^{+} \hat{j}_{\nu} \hat{G}^{-}
\right>_{\! c}
\: - \:
\left<
\hat{O}_{\mu} \hat{G}^{+}
\right>_{\! c}
\: 
\left<
\hat{j}_{\nu} \hat{G}^{-}
\right>_{\! c}
\; .
\label{eq-vertex}
\end{equation}
When dealing with transport properties, the vertex corrections are
crucial to discuss scattering processes at impurities as well as for
concentrated alloys
\cite{CB01,Sin08,SVW+15}.

To study how the treatment of the disorder in the host influences
calculated transport properties, we additionally did some calculations
for which the host was treated as a crystal with artificial atoms, as
in the virtual crystal approximation (VCA).  Specifically, the host
potential was constructed as a weighted average of single-site
potentials for Fe and Ni (taken from a CPA calculation for undoped Py)
and the corresponding atomic number was taken as a weighted average of
Fe and Ni atomic numbers.  Permalloy doped with V, Co, Pt, or Au was
then treated as in a non self-consistent CPA calculation, where for
the host potential we took the potential of the artificial ``VCA
permalloy crystal'' and the dopant potential was taken from a proper
CPA calculation.  Such a procedure does not represent a good
approximation for calculating transport properties, nevertheless, it
enables us to highlight the differences arising from treating the host
either as a periodic crystal or as a disordered alloy.


\subsection{Dealing with finite temperature effects}

\label{sec-aam}

Finite temperature effects were included by means of the so-called
alloy analogy model \cite{EMC+15}: temperature-induced atomic
displacements and spin fluctuations are treated as localized and
uncorrelated, giving rise to two additional types of disorder that can
be described using the CPA. For this approach, the atomic potentials are considered as
frozen. Local magnetic moments are assumed to be rigid, i.e., only
transversal fluctuations are taken into account.

Atomic vibrations were described using 14 displacement vectors, each
of them being assigned the same probability.  The lengths of these
displacement vectors were set to reproduce the temperature-dependent
root mean square displacement \mm{\sqrt{\langle u^{2} \rangle}}\ as
given by the Debye's theory \cite{EMC+15}.  Displacements for
different element types on the same site were taken as identical.  The
Debye temperature $\Theta_{D}$ was estimated for each composition as a
weighted average of Debye temperatures of the constituting elements.
Elemental Debye temperatures we used are listed in the Appendix.

Spin fluctuations were described by assuming that the local moments
are oriented along pre-defined vectors $\hat{\bm{e}}_{f}$ which are
isotropically distributed; we allowed for 60~values of the polar angle
$\theta_{f}$ and 3~values of the azimuthal angle $\phi_{f}$.  The
probability $x_{f}$ of each spin orientation was obtained by relying
on the mean-field theory \cite{EMC+15}.  Analogously to the treatment
of the displacements, the probabilities $x_{f}$ are taken 
independent on which element occupies a given site.

Setting the probability $x_{f}$ requires knowing the
temperature-dependent Weiss field parameter $w(T)$ \cite{EMC+15}.  It
can be obtained within the mean field theory if one knows the reduced
magnetization \mm{M(T)/M(0)}, i.e., the ratio of the magnetization
$M(T)$ to the magnetization at $T$=0~K (see Ref.~\onlinecite{EMC+15}
for more details).  The reduced magnetization \mm{M(T)/M(0)}\ is thus
an external input parameter for our calculation.  We assumed a
modified Bloch form for $M(T)$ according to
\begin{equation}
M(T) \: = \: M(0) \,
\left[
  1 \, - \, A \left( \frac{T}{T_{C}} \right)^{(3/2)}
\right]
\; ,
\label{eq-M-T}
\end{equation}
where $A$ is a constant.  The Curie temperature $T_{C}$ for undoped Py
can be taken from experiment (865~K).  For doped Py we first evaluated
the mean-field $T_{C}$ using exchange coupling constants obtained via
the Liechtenstein formula \cite{LKAG87,PKT+01,SME+19} and then
subtracted from it the difference between the mean-field and
experimental $T_{C}$ for undoped Py (this difference is 220~K in our
case).  This approach is consistent with the study of Devonport
\ea\ \cite{DVS+18} on Cr-doped Py, where a more-or-less constant
downward shift between theoretical and experimental $T_{C}$ can be
observed.  Finally, we set $A$=0.35 because this leads to a good
agreement between the model and experimental \mm{M(T)/M(0)}\ curves
for Pt-doped Py \cite{HGS+16}.  A representative selection of the
\mm{M(T)/M(0)}\ curves obtained thereby is presented in the Appendix.


\subsection{Technical details}

The real space representation of the Green function operator
\mm{\hat{G}}\ was evaluated within the ab-initio framework of
spin-density functional theory, relying on the generalized gradient
approximation using the Perdew, Burke and Ernzerhof (PBE)
functional \cite{PBE96}. The electronic structure embodied in the underlying
effective single-particle Dirac Hamiltonian was calculated in a fully
relativistic mode using the spin-polarized KKR-GF formalism
\cite{EKM11} as implemented in the {\sc sprkkr} code
\cite{sprkkr-code}.  For the multipole expansion of the Green
function, an angular momentum cutoff \mm{\ell_{\mathrm{max}}}=3 was
used.  The potentials were subject to the atomic sphere approximation
(ASA).  Self-consistent potentials were obtained employing energy
integration on a semicircle in a complex plane using 32 points, the
$\bm{k}$-space integration was carried out via sampling on a regular
mesh corresponding to $30^{3}$ $\bm{k}$-points in the full Brillouin
zone (BZ).

For the V, Pt, and Au dopants, the equilibrium lattice constant
$a_{0}$ was determined for each dopant concentration by minimizing the
total energy.  For the Co-doped Py, the conductivities were calculated
always for the lattice constant obtained for undoped Py because the
variation of $a_{0}$ with the Co concentration is very small and thus
hardly discernible from the numerical noise.  This is not surprising
given the mutual positions of Fe, Co, and Ni in the periodic table.

The Kubo-Bastin formulae Eqs.~(\ref{eq-Bastin})--(\ref{eq-Bastin-II})
were evaluated using similar settings as used for self-consistent
potentials except for the $\bm{k}$-space integrations for energy
points close to the real axis at \ef, where a very dense mesh has to
be used.  The choice of the $\bm{k}$-mesh is especially crucial for
crystalline hosts~\cite{KCE15} whereas if the host is an alloy the
integrands are smoother and performing the $\bm{k}$-mesh integration
is not so difficult.  Normally we used $576^{3}$ $\bm{k}$-points in
the full BZ at \ef\ and $288^{3}$ $\bm{k}$-points at the energy point
next-nearest to \ef.  In case of zero-temperature calculations for
undoped Py and for Py doped with Co, where the smoothening effect of
alloying may not be so strong, we used $1263^{3}$ $\bm{k}$-points at
\ef\ and $631^{3}$ $\bm{k}$-points at the energy point next-nearest to
\ef, to be on the safe side.



\section{Results and discussion}   \label{sec-res}


\subsection{Longitudinal conductivity and AMR}   

\label{sec-sigxx}


\subsubsection{Calculation of transport quantities}   

\label{sec-transcal}

\begin{SCfigure*}
\includegraphics[viewport=0.4cm 0.5cm 12.4cm 13.5cm]{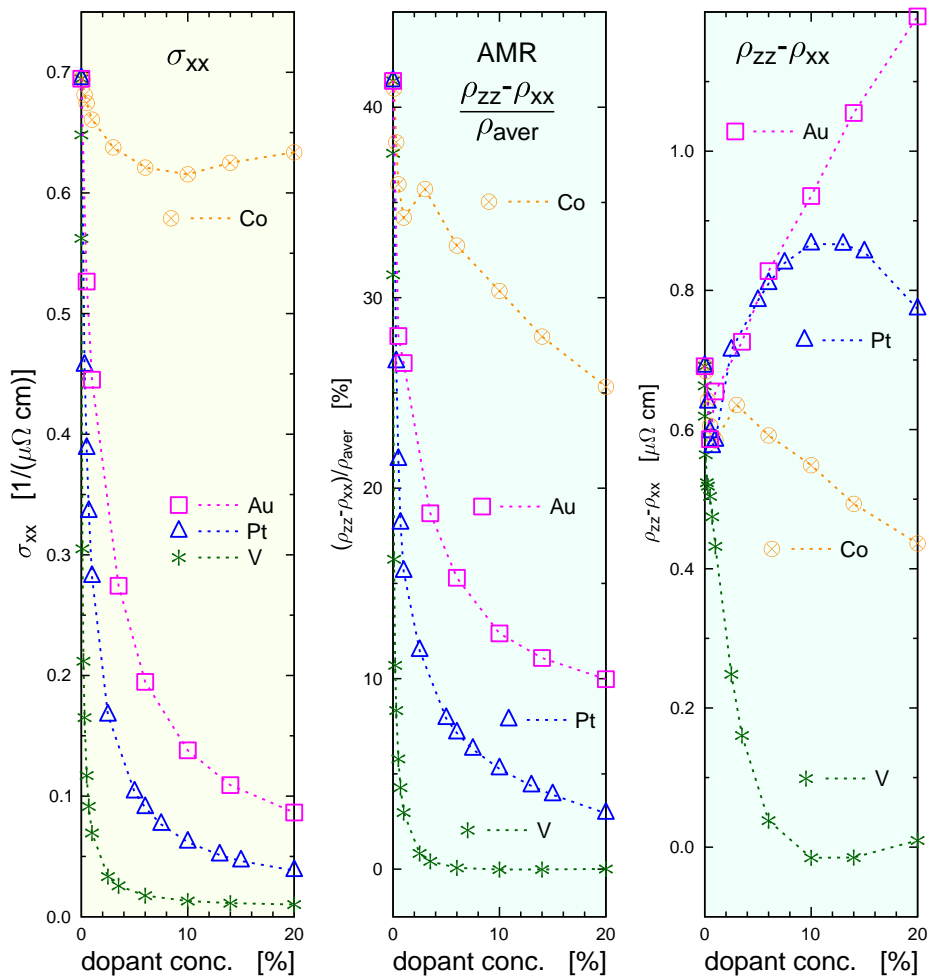}%
\caption{(Color online) Theoretical longitudinal conductivity
  $\sigma_{xx}$ (left), relative anisotropic magnetoresistance
  $(\rho_{zz}-\rho_{xx})/\rho_{\text{aver}}$ (middle), and the
  difference of resistivities $\rho_{zz}-\rho_{xx}$ (right) for Py
  doped with V, Co, Pt, and Au, for $T$=0~K.  Calculated values are
  shown by markers, the lines are guides for an eye. The dopant type
  is indicated in the legend.}
\label{fig-sigxx-T0}
\end{SCfigure*}

Doping of Py leads to a decrease of the longitudinal charge
conductivity \sigxx, as illustrated in Fig.~\ref{fig-sigxx-T0} (left).
The rate of the decrease depends on the dopant, following the sequence
Co--Au--Pt--V.  The same sequence characterizes also the dependence of
the anisotropic magnetoresistance on the dopant concentration (middle
panel of Fig.~\ref{fig-sigxx-T0}).  This is, to a large extent, due to
the definition of AMR as
\begin{displaymath}
\frac{\rho_{zz}-\rho_{xx}} {\rho_{\text{aver}}}
\end{displaymath}
because it is normalized to the average resistance
\begin{equation}
  \rho_{\text{aver}} \: = \: \frac{1}{3}
  \left[
    2 \rho_{xx} \, + \,  \rho_{zz}
    \right]
  \; .
\end{equation}
To provide an uncompensated picture, we show the bare difference
\mm{\rho_{zz}-\rho_{xx}}\ in the right panel of
Fig.~\ref{fig-sigxx-T0}.

We calculated also spin-resolved conductivities, following the
prescription
\begin{eqnarray}
  \sigma_{xx}^{\text{(maj)}} & = & \frac{1}{2} \,
  \left( \sigma_{xx} \, + \,  \sigma_{xx}^{z} \right) \; ,
  \label{eq-sigmaj}  \\
  \sigma_{xx}^{\text{(min)}} & = & \frac{1}{2} \,
  \left( \sigma_{xx} \, - \,  \sigma_{xx}^{z} \right) 
  \quad .
    \label{eq-sigmin} 
\end{eqnarray}
We found that practically all the transport is mediated by
majority-spin electrons: $\sigma_{xx}^{\text{(min)}}$ is by an order
of magnitude smaller than $\sigma_{xx}^{\text{(maj)}}$ (results not
shown).  In a two-current model, this would imply a large AMR --- see
Eq.~(6) in Campbell \ea\ \cite{CFJ70}.

\begin{SCfigure*}
\includegraphics[viewport=0.5cm 0.5cm 13.0cm 18.0cm]{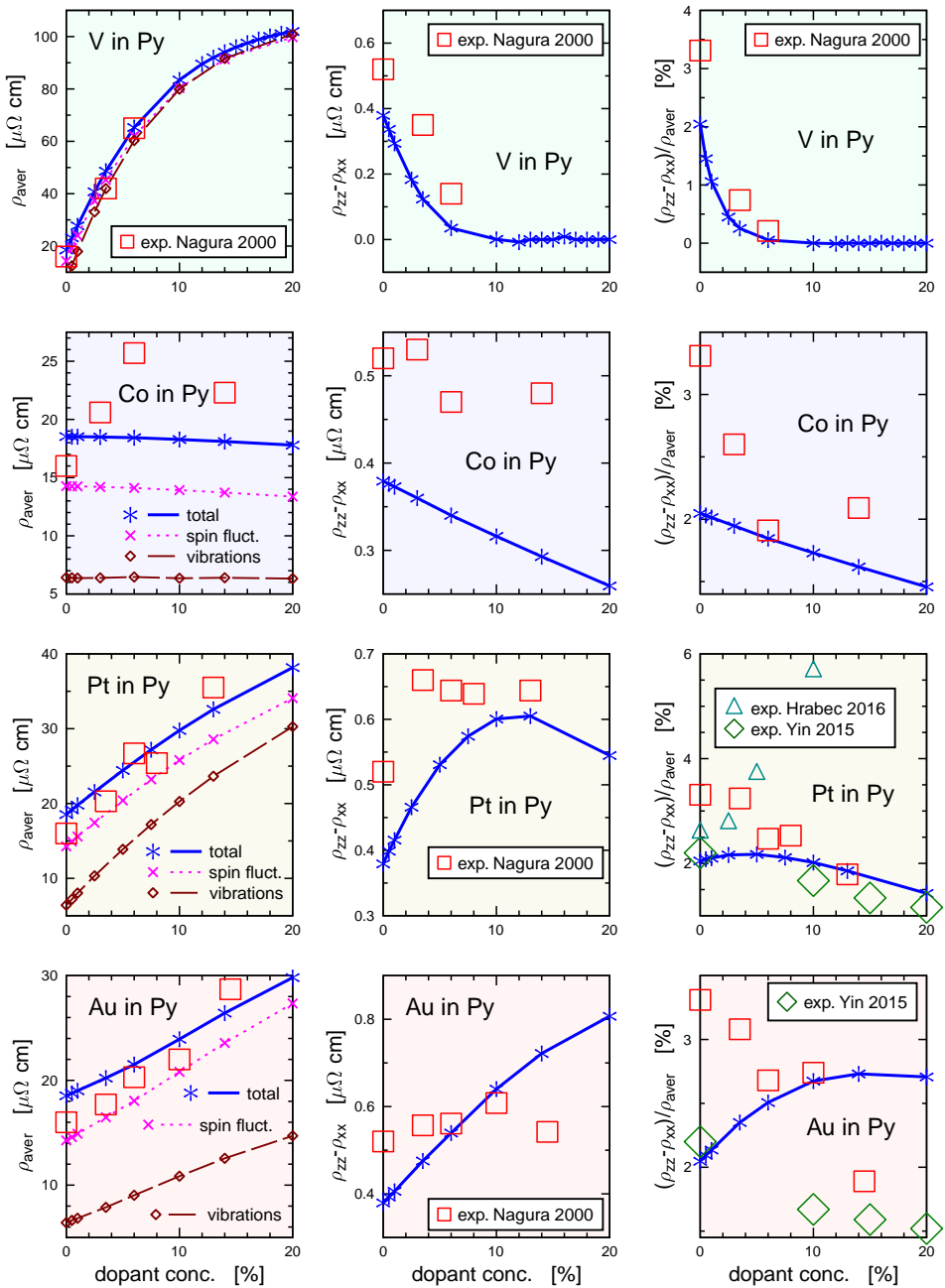}%
\caption{(Color online) Average resistivity \rhoav, difference between
  resistivities \mm{\rho_{zz}-\rho_{xx}}, and anomalous
  magnetoresistance $(\rho_{zz}-\rho_{xx})/\rho_{\text{aver}}$ for Py
  doped with V, Co, Pt, and Au calculated for $T$=300~K, compared to
  experiment.  Experimental data are from Nagura \ea\ \protect\cite{NST+00},
  Yin \ea\ \protect\cite{YPA+15}, and Hrabec
  \ea\ \protect\cite{HGS+16}.  For the 
  average resistivity \rhoav, theoretical results obtained when only
  the spin fluctuations or only the atomic vibrations are accounted
  for, are also shown.}
\label{fig-rho-T300}
\end{SCfigure*}

Experimental data are available only for room temperature.  Therefore
we present in Fig.~\ref{fig-rho-T300} theoretical results for the
average resistivity \mm{\rho_{\text{aver}}}, for the difference
between resistivities \mm{\rho_{zz}-\rho_{xx}}, and for the
anisotropic magnetoresistance
$(\rho_{zz}-\rho_{xx})/\rho_{\text{aver}}$ obtained via the alloy
analogy model for $T$=300~K, together with available experimental
data.  There is a good overall agreement between theory and
experiment
of Nagura \ea\ \cite{NST+00} as concerns the average
resistivity \mm{\rho_{\text{aver}}}.  The agreement is less good for
the AMR; the trends are mostly described correctly but there are
differences in the absolute values.
For Au-doped Py, however, the trends of experimental and theoretical AMR do not
agree (middle bottom and right bottom panels in
Fig.~\ref{fig-rho-T300}).
The experimental AMR data for Pt-doped Py of Hrabec
\ea\ \cite{HGS+16} exhibit a different trend than the experimental
data of Nagura \ea\ \cite{NST+00} and of Yin \ea\ \cite{YPA+15}, as
well as than our theoretical data; the reason for this is unclear.

When assessing the agreement and disagreement between theory and
  experiment in Fig.~\ref{fig-rho-T300}, one should keep in mind that
  the experimental data may also be affected by uncertainties, e.g.,
  concerning the concentration of the dopants in the sample or the
  presence of other defects.  Note also that the AMR values obtained
  by different experiments differ approximately by the same amount as
  theory and experiment.  We conclude that, as a whole, our
  theoretical data agree with the experimental data presented in
  Fig.~\ref{fig-rho-T300} quantitatively as concerns
  \mm{\rho_{\text{aver}}} and qualitatively as concerns the AMR.  We
  assume that this gives reasonable confidence that our theoretical
  results can be relied on as concerns the general trends and
  tendencies.

\begin{figure}
\includegraphics[viewport=0.5cm 0.5cm 6.0cm 5.0cm]{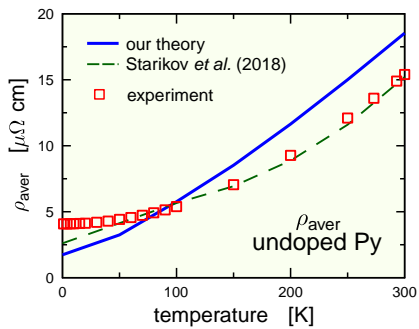}%
\caption{(Color online) Dependence of the resistivity
  $\rho_{\text{aver}}$ of undoped Py on the temperature as provided by
  our calculation (blue solid line), by calculations of Starikov
  \ea\ \protect\cite{SLY+18} (green dashed line), and by experiment
  \protect\cite{HAW+83} (red squares).  }
\label{fig-rho-temp}
\end{figure}

Another test of our calculations can be done for undoped Py by
comparing the temperature-dependence of the calculated resistivity
\rhoav\ with available experimental data \cite{HAW+83}.  This is done
in Fig.~\ref{fig-rho-temp}, where the experimental data are shown
together with our results and also with calculations of Starikov
\ea\ \cite{SLY+18}, who modeled the temperature-induced disorder by
means of supercells.  Our calculation of \rhoav\ accounts quite
  well for the trend but the agreement with experiment is less good
  than for the calculations of Starikov \ea\ \cite{SLY+18}.  A
  possible reason for this may be the different models used to
  describe the thermal disorder by Starikov \ea\ \cite{SLY+18}
  (supercells) and by us (CPA,  as outlined in Ebert
   \cite{EMC+15}).

To provide a more specific view
on the impact of temperature-induced disorder, 
data for \rhoav\ are shown not only
for the case when spin fluctuations and atomic vibrations are included
together but also when either only the spin fluctuations are included
or only the atomic vibrations are included.  Note that for the
V~dopant, the three data sets are hardly distinguishable from each
other on the scale of Fig.~\ref{fig-rho-T300}.
The contribution to the resistivity due to thermal spin
  fluctuations,
 \[
  \rho_\text{aver}^{\text{(sfluct)}}(300~\text{K}) -
  \rho_\text{aver}(0) \; ,
  \] 
 is typically about three times larger than the contribution due to
atomic vibrations,
 \[
  \rho_\text{aver}^{\text{(vibr)}}(300~\text{K}) -
  \rho_\text{aver}(0) \; .
  \] 
An exception is the case of V as dopant, for which spin
  fluctuations dominate for low V concentrations whereas for
  concentrations larger than about 5~\% we found that both
  contributions are comparable.  (This last fact is not discernible on
  the scale of Fig.~\ref{fig-rho-T300}.) 

We checked also the  applicability of the 
  Matthiessen rule, i.e., whether the
  influence of atomic vibrations and of spin fluctuations is additive.
  The equation
\begin{equation}
[\rho_\text{aver}^{\text{(sfluct)}}(T) -
  \rho_\text{aver}(0)]
\: + \:
[\rho_\text{aver}^{\text{(vibr)}}(T) -
  \rho_\text{aver}(0)]
\; = \;
\rho_\text{aver}^{\text{(combi)}}(T) -
\rho_\text{aver}(0)
\; ,
\label{eq-matt}
\end{equation}
with $\rho_\text{aver}^{\text{(combi)}}(T)$ denoting the resistivity
when both spin fluctuations and atomic vibrations are accounted for
simultaneously, is always satisfied with an accuracy better than 5~\%
(typically about 1~\%).  We checked that this is true also for other
temperatures (between 0~K and 300~K).  The breakdown of the
Matthiessen rule observed for some other systems
\cite{GPB+14,EMC+15,DKW+19} thus does not occur here.


\subsubsection{Electronic structure analysis}   

\label{sec-bsf}

\begin{figure}
\includegraphics[viewport=0.5cm 0.5cm 9.0cm 10.5cm]{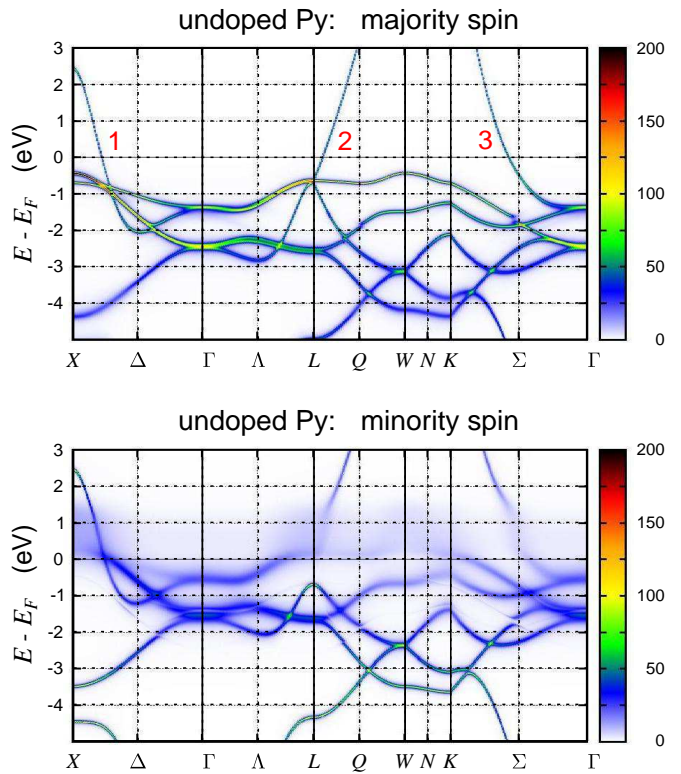}%
\caption{(Color online) Bloch spectral function $A_{B}(\bm{k},E)$ of
  undoped Py projected on majority-spin states and on minority-spin
  states.  Three instances where the majority-spin band crosses the
  Fermi level are marked by the numbers 1, 2, and 3.}
\label{fig-bloch}
\end{figure}

To get an intuitive insight into the sequence Co--Au--Pt--V which
characterizes the efficiency of various dopants in suppressing the
high conductivity of Py (see Fig.~\ref{fig-sigxx-T0}), we inspect how
the doping affects relevant aspects of the electronic structure.
First, the Bloch spectral function $A_{B}(\bm{k},E)$ of undoped Py
projected on majority-spin and minority-spin states, respectively, is
shown in Fig.~\ref{fig-bloch}.  As it is well-known, the majority-spin
states form well-defined bands, demonstrating that the disorder is
weak for these states.  For minority-spin states, on the other hand, a
significant smearing of the bands is evident.

\begin{SCfigure*}
\includegraphics[viewport=0.5cm 0.5cm 12.5cm 6.3cm]{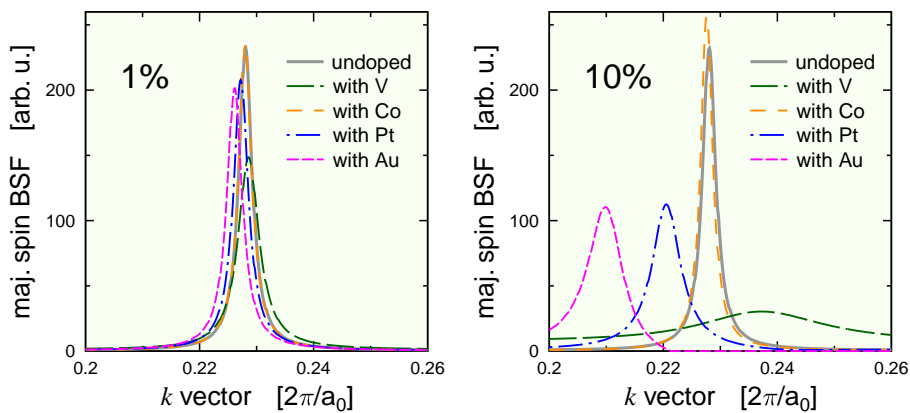}%
\caption{(Color online) Majority-spin Bloch spectral function at the
  Fermi level $A_{B}(\bm{k},E_{F})$ for $\bm{k}$ in the vicinity of
  the point marked as 1 in Fig.~\ref{fig-bloch}.  The $\bm{k}$~vector
  lies on the $X$--$\Delta$ path, its values are in units of
  \mm{2\pi/a_{0}}.  The concentration of the dopants is 1\% (left) and
  10\% (right).}
\label{fig-fwhm}
\end{SCfigure*}

Doping Py with V, Co, Pt, or Au introduces smearing for the
majority-spin states as well.  From the point of view of transport
properties, the most important changes occur around \ef.  A detailed
view how the doping influences the Bloch spectral function at \ef\ is
presented in Fig.~\ref{fig-fwhm}.  Here, we show majority-spin Bloch
spectral function $A_{B}(\bm{k},E_{F})$ for $\bm{k}$ in the vicinity
of the point marked as 1 in Fig.~\ref{fig-bloch}, for Py doped with
1\% and 10\% of V, Co, Pt, and Au.  For the 1\% dopant concentration,
the changes with respect to the undoped case are relatively small, as
to be expected.  For the 10\% dopant concentration, the changes are
obvious.  Note that as the spin is not a good quantum number (because
of the SOC), the projection of $A_{B}(\bm{k},E_{F})$ on the spin
directions cannot be done unambigously. Namely, it reflects not only
the exchange coupling but also the hybridization of spin states.
However, our analysis is not really hindered by this.

\begin{table*}
\caption{FWHM's of peaks (in units of \mm{2\pi/a_{0}}) of
  majority-spin Bloch spectral function $A_{B}(\bm{k},E_{F})$ for
  doped Py if the $\bm{k}$~vector is in the vicinity of points marked
  as 1, 2, and 3 in Fig.~\ref{fig-bloch}.  Concentration of the
  dopants is 1\% and 10\%.}
\label{tab-fwhm}
\begin{ruledtabular}
\begin{tabular}{ldddddd}
  \multicolumn{1}{c}{} &
  \multicolumn{2}{c}{$\bm{k}$ point 1} &
  \multicolumn{2}{c}{$\bm{k}$ point 2} &
  \multicolumn{2}{c}{$\bm{k}$ point 3} \\
  \multicolumn{1}{c}{ } &
  \multicolumn{1}{c}{1\%} &
  \multicolumn{1}{c}{10\%} &
  \multicolumn{1}{c}{1\%} &
  \multicolumn{1}{c}{10\%} &
  \multicolumn{1}{c}{1\%} &
  \multicolumn{1}{c}{10\%}  \\
\hline
undoped &  \multicolumn{2}{c}{0.0028}  &
\multicolumn{2}{c}{0.0032} &
\multicolumn{2}{c}{0.0039} \\
Co  &  0.0028  &  0.0025  &  0.0032  &  0.0031 &  0.0039  &  0.0038 \\
Au  &  0.0033  &  0.0068  &  0.0034  &  0.0054 &  0.0040  &  0.0063 \\
Pt  &  0.0032  &  0.0061  &  0.0036  &  0.0074 &  0.0046  &  0.0121 \\
V   &  0.0045  &  0.0461  &  0.0068  &  0.0611 &  0.0118  &  0.1556    
\end{tabular}
\end{ruledtabular}
\end{table*}

One can be quantitative and evaluate for each of the peaks in
$A_{B}(\bm{k},E_{F})$ the corresponding full width at half maximum
(FWHM).  We performed this for all three bands which cross the Fermi
level in Fig.~\ref{fig-bloch}; the results are shown in
Table~\ref{tab-fwhm}.  Generally, the FWHM's increase if the dopant is
varied along the sequence Co--Au--Pt--V.  The exception to this rule
is the Au-Pt pair in the vicinity of the $\bm{k}$~point~1; this
deviation is probably due to the SOC.

The FWMH of the Bloch spectral function corresponds to the inverse of
the electronic life-time that enters the semiclassical theory of
electron transport.  We also checked the effect of the doping on the
slope of the energy band which is linked to the corresponding group
velocity.  These variations are only few percents (even for 10\%
doping), with no clear systematic trend.  The density of states at
\ef, which could be linked to the number of carriers, does not exhibit
a systematic trend either and the variations with the doping are less
than 10\% (corresponding data are not shown).  We conclude, therefore,
that the decrease of the conductivity of Py upon doping can be traced,
first of all, to the decrease of the electron life-time.

\begin{figure}
\includegraphics[viewport=0.5cm 0.5cm 9.0cm 10.0cm]{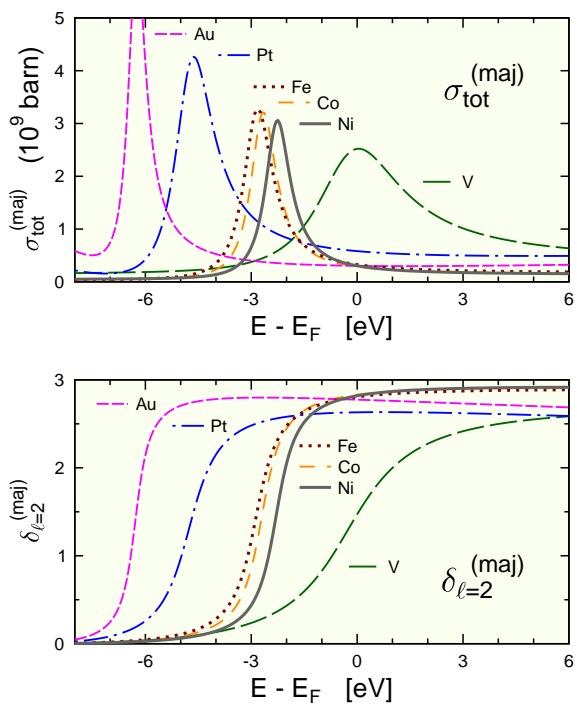}%
\caption{(Color online)  Lower panel: Atomic phase-shifts of majority-spin
  $d$~electrons for scattering at Fe and Ni as host atoms and at V, Co,
  Pt, and Au as impurity atoms in Py.  Upper panel: 
  Total atomic scattering
  cross-section of majority-spin electrons for the same types. }
\label{fig-phase}
\end{figure}

For further insight, we analyze in the following the scattering
  of majority-spin electrons. As the current mediated by minority-spin
  electrons was found to be practically negligible --- see the text
  below Eq.~(\ref{eq-sigmin}) --- we ignore it in this analysis.  The
  lower panel of Fig.~\ref{fig-phase} depicts the phase-shifts of the
  majority-spin $d$ electrons (which dominate around \ef) for Fe and
  Ni host atoms and for V, Co, Pt, and Au impurities.  The results we
  show here were obtained for  nominally zero concentrations of
  the dopants, 
  i.e., for the single-site impurity limit but a similar picture would
  arise also for finite doping concentrations.  One can see that the
  scattering properties of Co are very similar to those of Fe and Ni
  in Py,  implying  that doping Py with Co will influence the transport
  properties only a little.  The influence of V, Pt, and Au, on the
  other hand, will be much more significant as their atomic scattering
  properties differ much more from those of the host atoms. 

 The total cross-section for scattering of majority-spin
  electrons $\sigma_{\text{tot}}^{\text(maj)}$ at each of the atoms
  (including all angular-momentum components up to $\ell$=3) is shown
  in the upper panel of Fig.~\ref{fig-phase}.  The most significant
  energy region for transport is around \ef.  The corresponding
  cross-section $\sigma_{\text{tot}}^{\text(maj)}(E_{F})$ decreases in
  the order V--Pt--Au, in line with the efficiency of these elements
  to reduce the conductivity of doped Py (cf.\ left panel of
  Fig.~\ref{fig-sigxx-T0}). 

 One can see this as an illustration that transport properties of
  doped systems (and compounds and alloys in general) stem from a
  delicate interplay of the electronic structure of the constituting
  elements.  Vanadium is a 3$d$ element, hence one would expect that
  its local electronic structure differs less from the electronic
  structure of Py than the local electronic structure of 5$d$ elements
  Pt and Au.  Consequently, one might assume that V impurities will
  present a smaller disturbance for transport in Py than Pt or Au
  impurities.  Nevertheless, the V dopants influence the conductivity
  of Py more than the Pt or Au dopants (Fig.~\ref{fig-sigxx-T0}).
  This is clearly because what matters most is the situation at \ef.
  Fig.~\ref{fig-phase} shows that the scattering cross-section at V
  atoms in Py has a peak just at \ef, whereas the cross-sections at Pt
  and Au atoms in Py have their maxima at lower energies.  For other
  combinations of host and impurity atoms, the situation might
  obviously be quite different. 
 Seen from another perspective, one can argue that the same
  mechanism which leads to strong scattering of electrons by V atoms
  in Py leads also to creation of virtual bound states above \ef\ for
  a V impurity in Ni \cite{SOZD87}.

 Finally, let us note that the efficiency of V atoms in reducing
  the conductivity of Py is {\em not} directly linked to the
  antiparallel orientation of magnetic moments of V atoms with respect
  to moments of host atoms.  To check this, we manipulated the local
  exchange field $B_{\text{ex}}$ of V atoms so that it is the same as
  the average of the exchange fields of the Fe and Ni atoms (and the
  magnetic moment of V atoms is  oriented parallel to the moment of the host
  atoms).  For this situation, the conductivity of V-doped permalloy
  changes typically by only about 20~\% and, accordingly, the overall
  picture as provided by Fig.~\ref{fig-sigxx-T0} remains essentially
  unchanged.



\subsection{Dependence of AHE and SHE on dopants 
  concentration for $T$=0~K}   

\label{sec-sigxy}

\begin{figure}
\includegraphics[viewport=0.5cm 0.5cm 9.0cm 13.5cm]{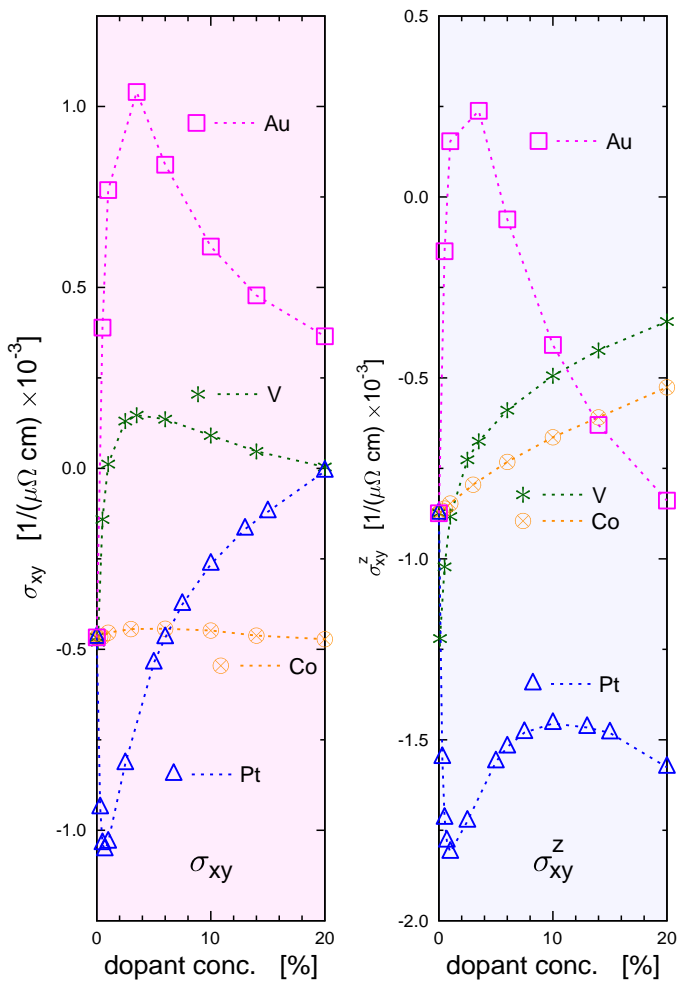}%
\caption{(Color online) Theoretical off-diagonal conductivities
  \sigxy\ (corresponding to AHE) and \sigZxy\ (corresponding to SHE)
  for Py doped with V, Co, Pt, and Au, for $T$=0~K.  Calculated values
  are shown by markers, the lines are guides for an eye. The dopant types
  are indicated in the legend.}
\label{fig-offconc}
\end{figure}

Off-diagonal conductivities \sigxy\ and \sigZxy\ calculated for
different dopant types are shown in Fig.~\ref{fig-offconc}.  The
dependence of \sigxy\ and \sigZxy\ on the dopant concentration is
highly non-monotonic.  Doping with Co introduces only small changes
with respect to the undoped case.  Doping with Pt and Au --- elements
with a strong SOC --- introduces large changes: even the sign of
\sigxy\ or \sigZxy\ can be reverted in this way.  
If the dopant
concentration approaches zero, values of \sigxy\ and \sigZxy\ smoothly
acquire the values which correspond to undoped Py.  Note that this is
true also for \sigZxy\ in the case of the V~dopant, even though it is
not clearly visible at the scale of Fig.~\ref{fig-offconc}: we
verified that \sigZxy\ has a very sharp minimum for the 0.1\%
V~concentration whereas lowering the doping level further leads to
values of \sigZxy\ which gradually approach the value of \sigZxy\ 
of undoped Py.
 A non-monotonic dependence of \sigxy\ and \sigZxy\ 
  on the dopant concentration appears to be a general feature for the
  systems we study.

\begin{figure*}
\includegraphics[viewport=0.5cm 0.5cm 17.7cm 13.5cm]{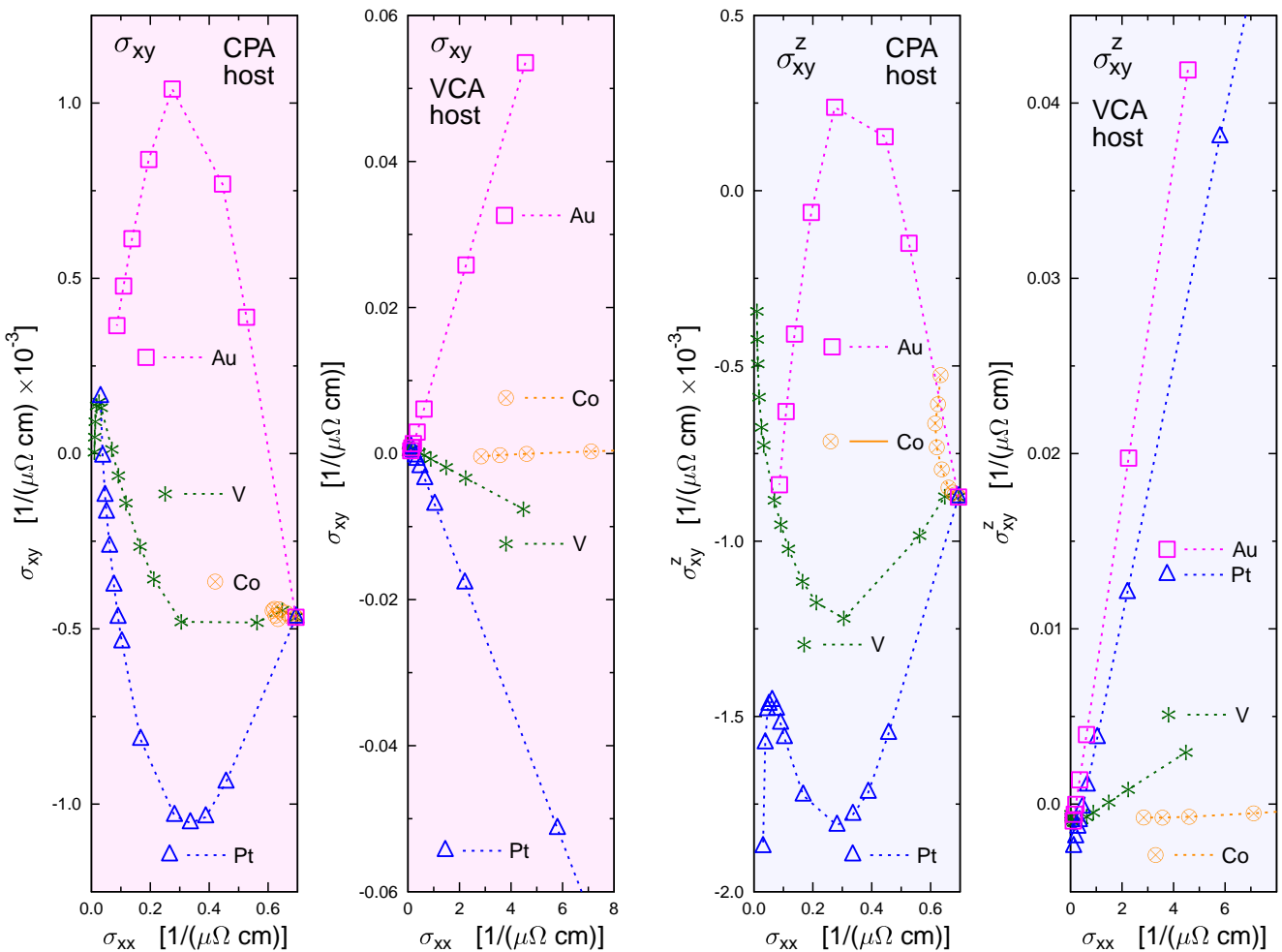}%
\caption{(Color online) Theoretical dependence of the AHE conductivity
  \sigxy\ (two left panels) and of the SHE conductivity \sigZxy\ (two
  right panels) on the longitudinal conductivity \sigxx\
if the concentration of the dopants is varied (for $T$=0~K).
  The Py host
  is treated either within the CPA or within the VCA, respectively, as
  indicated at the top of the panels. The dopant type is specified in
  the legend. }
\label{fig-xyonxx}
\end{figure*}

It has
been well-established that \sigxy\ and \sigZxy\ diverge in the clean
limit \cite{OSN+08,NSO+10,SVW+15,HSK+16}.  However, in our case we are
not in the clean limit even for zero dopant concentration because the
host is not a crystalline metal but a substitutional alloy, with a
finite longitudinal conductivity \sigxx.
Consequently, neither \sigxy\ or \sigZxy\ diverge at low dopants
concentrations, in 
contrast to the situation for crystalline hosts.  

Another feature which highlights the difference between crystalline
and disordered host is that,
if the dopant concentration is varied, 
the off-diagonal
conductivities \sigxy\ and \sigZxy\ are not proportional to the
longitudinal conductivity \sigxx\ for low dopant concentrations,
unlike what is common for crystalline hosts
\cite{OSN+08,LGK+11,CFH+15}.  This can be clearly seen in
Fig.~\ref{fig-xyonxx}.  If the host is treated
 as a truly disordered system, i.e., 
within the CPA, the
dependence of \sigxy\ and \sigZxy\ on \sigxx\ is complicated and
non-monotonous.    Note that different
dopants give rise to quite different dependencies of \sigxy\ or
\sigZxy\ on \sigxx.

The fact that the host is a disordered system means,
 among others,
that
the dependence of  the AHE and SHE on the dopant concentration
  cannot be described unambigously 
in terms of skew scattering, side-jump
scattering, or intrinsic contribution in the same way as 
 it can be done when investigating the effect of doping for a
crystalline host \cite{CB01,OSN+08,LGK+11,CFH+15,HSK+16}. 
The above mentioned scheme, namely, assumes that for zero
dopant concentration the electron participating in the transport is
not scattered.  This is true only if the host is a perfect periodic
crystal.
 If the host material is an alloy, the concepts of skew
  scattering and side-jump scattering can be misleading. 

In our theoretical scheme, having a perfect crystal for a host would
correspond to treating undoped Py not within the CPA but within the VCA.
To illustrate the effect of the host disorder more clearly, we present
in Fig.~\ref{fig-xyonxx} the dependence of \sigxy\ and \sigZxy\ on
\sigxx\ also if the host is treated within the VCA.  One can see that
in this case both \sigxy\ and \sigZxy\ depend  linearly on \sigxx\
if the dopant concentration is low (i.e., the \sigxx\ conductivity is
high).  The host disorder thus has a crucial role in the non-monotonic
dependence of \sigxy\ and \sigZxy\ on the dopant concentration
(cf.\ Fig.~\ref{fig-offconc}).

 It should be noted that the conclusion that the concepts of skew
  and side-jump scattering are not directly applicable to the analysis
  of the AHE and SHE in disordered hosts concerns specifically the
  dependence of \sigxy\ and \sigZxy\ on the concentration of the
  dopants.  Earlier studies dealing with the mechanism of AHE in fully
  or partially disordered systems were concerned with the dependence
  of AHE either on the thickness of thin film samples
  \protect\cite{SLH+14,LHY+15} or on the degree of partial order
  \protect\cite{ZPZ+14,HSK+16}.  More sophisticated frameworks
  distinguishing more sources of scattering
  \protect\cite{HST+15,YJ+17} might be, in principle, applicable to
  our systems but probably not in a straightforward way: this can be
  seen, e.g., from the fact that in our case the AHE conductivity
  \sigxy\ is {\em not} linearly proportional to \sigxx\ at $T$=0~K,
  contrary to what formed the basis for earlier multivariable-scaling
  analyses --- cf.\ Eq.~(10) of Yue and Jin \cite{YJ+17}.  A different
  approach proposed by Bianco \ea\ \cite{BRS+14} uses supercells to
  describe the disorder and focuses on how it affects the Berry-phase
  contribution to the AHE, evaluated formally as for an ordered
  crystal.  Despite its conceptual appeal, this approach seems
  difficult to apply for impurity concentrations of just few percents
  because of technical issues linked to dealing with large
  supercells. 

 Similarly as we did in the case of the longitudinal transport,
  we also checked the effect of the exchange field $B_{\text{ex}}$ on
  the results.  We found that  if the local exchange field
$B_{\text{ex}}$ of the dopant atoms is manipulated so that it is equal
to the average of exchange fields of the Fe and Ni host atoms, no
significant changes in the calculated values of \sigxy\ or
\sigZxy\ occur (data not shown).   The antiparallel orientation
  of magnetic moment of the V atoms with respect to the host is thus
  not crucial for longitudinal as well as transverse transport
  properties of V-doped permalloy.



\subsection{Dependence of AHE and SHE on the temperature}   

\label{sec-temp}

\begin{figure}
\includegraphics[viewport=0.5cm 0.5cm 9.0cm 13.5cm]{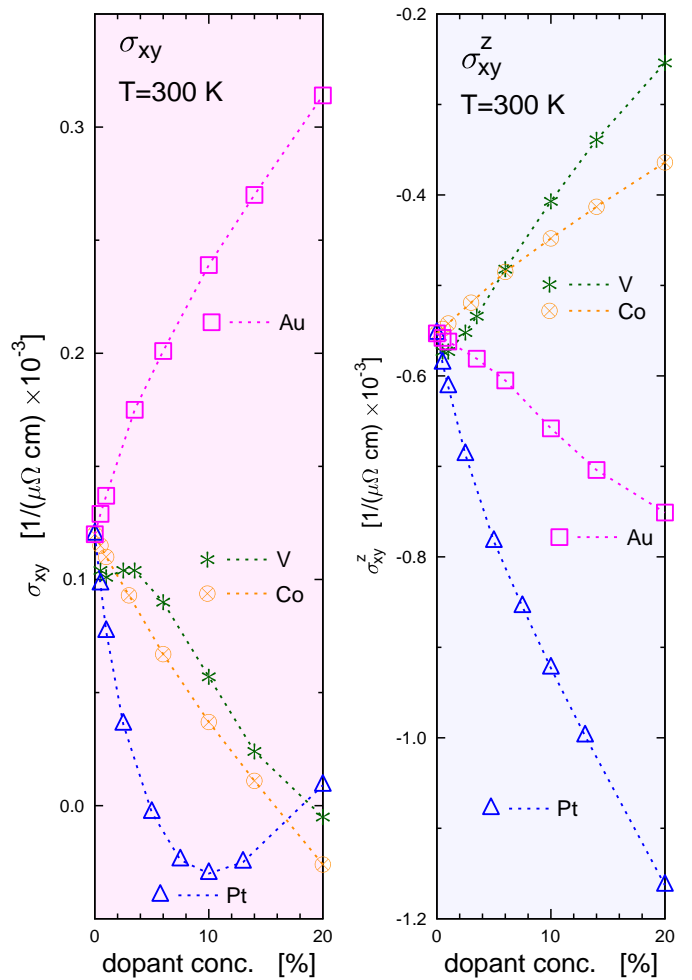}%
\caption{(Color online)  Theoretical off-diagonal conductivities
  \sigxy\ (corresponding to AHE) and \sigZxy\ (corresponding to SHE)
  for Py doped with V, Co, Pt, and Au, for $T$=300~K. }
\label{fig-conc-T300}
\end{figure}

Experiments dealing with electron transport are often done at room
temperature.  Therefore we present in Fig.~\ref{fig-conc-T300} the
dependence of \sigxy\ and \sigZxy\ on the concentration of V, Co, Pt,
and Au dopants for $T$=300~K, as it follows from the alloy analogy
model. This plot should be compared with Fig.~\ref{fig-offconc} where
the same dependence is explored for $T$=0~K.  It can be seen that the
effect of the temperature is really significant: the dependence of
\sigxy\ and \sigZxy\ on the dopants concentration is quite different
for $T$=0~K and for $T$=300~K.

\begin{figure}
\includegraphics[viewport=0.5cm 0.5cm 8.5cm 9.5cm]{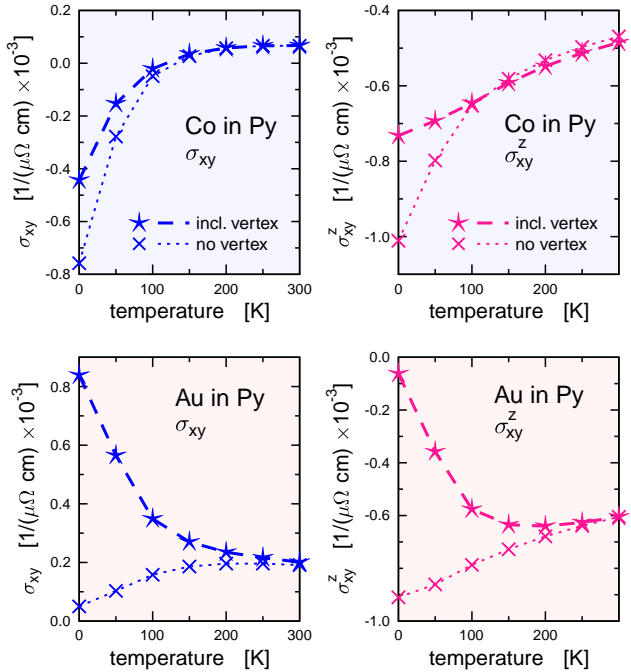}%
\caption{(Color online) Temperature-dependence of \sigxy\ and
  \sigZxy\ for Py doped with 6\% of Co and with 6\% of Au, calculated
  for the vertex corrections either included or omitted.}
\label{fig-temp-VC}
\end{figure}

A possible reason why the concentration dependencies of \sigxy\ and
\sigZxy\ change so much with temperature is that the vertex
corrections
 [see Eq.~(\ref{eq-vertex}) and the accompanying text in
    Sec.~\ref{sec-kubo}]   
become less important when the temperature 
increases.
 Intuitively this can be viewed that, 
 in a semi-classical picture, an electron
undergoes many scattering events if the temperature is high; there
is more disorder, majority-spin electron states loose their
crystal-like character, and the differences between various
trajectories effectively decrease.  All electrons undergo same
scattering events in the end, albeit in a different sequence, and the
vertex corrections become unimportant.  To illustrate this point, we
present in Fig.~\ref{fig-temp-VC} the temperature-dependence of
\sigxy\ and \sigZxy\ for Py doped with 6\% of Co and 6\% of Au
calculated with the vertex corrections either included or omitted.
For large enough temperatures the effect of vertex corrections is
getting negligible.

As concerns the longitudinal conductivity \sigxx, we checked that the
vertex corrections are not important even for zero temperature.  In
accordance with this, the overall pattern characterizing the
dependence of \sigxx\ on the concentration of various dopants for
$T$=0~K (Fig.~\ref{fig-sigxx-T0}) does not change if the temperature
increases, even though the numerical values of \sigxx\ obviously
decrease if the temperature rises (results not shown).

\begin{figure}
\includegraphics[viewport=0.5cm 0.5cm 8.5cm 5.0cm]{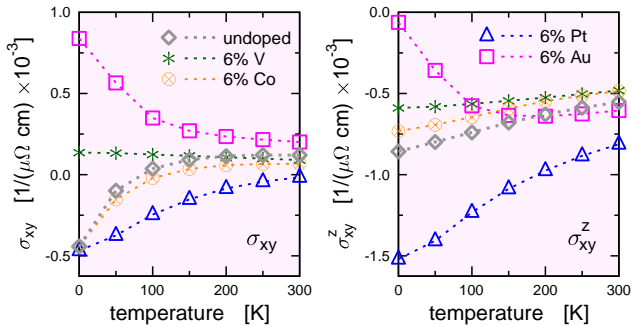}%
\caption{(Color online) Theoretical temperature-dependence of
  \sigxy\ and \sigZxy\ for undoped Py (diamonds) and for Py doped with
  6\% of V (asterisks), Co (crossed circles), Pt (triangles), and Au
  (squares). }
\label{fig-temp-types}
\end{figure}

If the temperature increases, the thermal effects should dominate and,
consequently, the differences between various dopings should decrease.
This can be seen in Fig.~\ref{fig-temp-types}, where we present the
temperature-dependence of \sigxy\ and \sigZxy\ for undoped Py as well
as for Py doped with 6\% of V, Co, Pt, and Au.

\begin{SCfigure*}
\includegraphics[viewport=0.5cm 0.5cm 12.6cm 9.5cm]{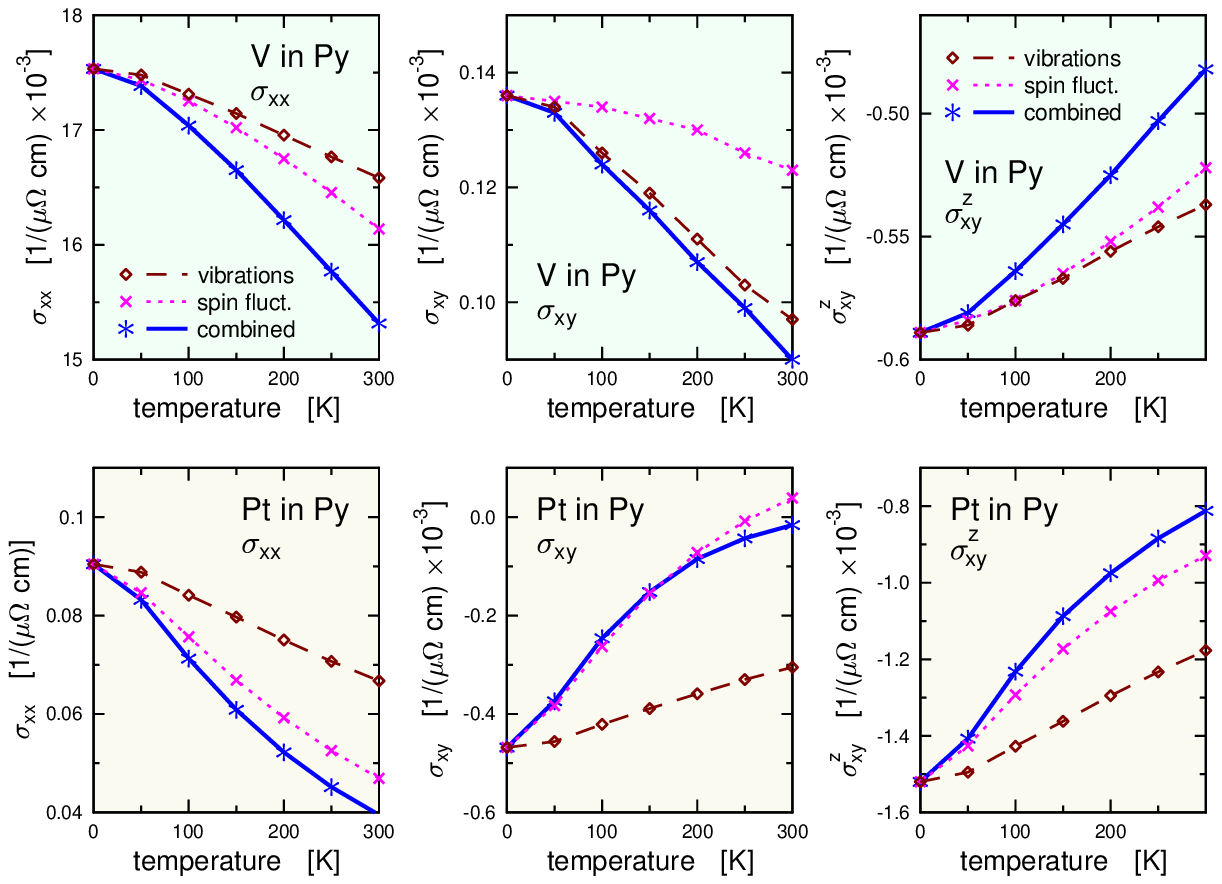}%
\caption{(Color online) Temperature-dependence of the longitudinal
  conductivity \sigxx\ and of off-diagonal conductivities \sigxy\ and
  \sigZxy\ for Py doped with 6\% of V (upper panels) and with 6\% of
  Pt (lower panels). Data are shown separately for calculations where
  both spin fluctuations and atomic vibrations are accounted for (full
  lines with asterisks), where only the spin fluctuations are
  considered (dotted lines with crosses), and where only the atomic
  vibrations are taken into account (dashed lines with diamonds).}
\label{fig-temp-parts}
\end{SCfigure*}

The alloy analogy model we employ takes into account atomic vibrations
and spin fluctuations together, on the same footing.  We can also
investigate both effects separately, as it was done for \rhoav\ in
Fig.~\ref{fig-rho-T300}.  Illustrative results concerning the
temperature-dependence of \sigxx, \sigxy, and \sigZxy\ for Py doped
with 6\% of V and 6\% of Pt are shown in Fig.~\ref{fig-temp-parts}.
It seems that there is no universal rule allowing to estimate
beforehand which of the two effects --- atomic vibrations or spin
fluctuations --- will be more important
 for the way the AHE or SHE are affected by the temperature. 
  E.g., the
temperature-dependence of \sigxy\ is dominated by atomic vibrations
for V-doped Py and by spin fluctuations for Pt-doped Py.
 We also checked whether the effects of atomic vibrations and of
  spin fluctuations on the AHE and SHE are additive, i.e., whether the
  Matthiessen rule Eq.~(\ref{eq-matt}) holds also for \rhoxy.  We
  found that for V, Co, and Au dopants it is satisfied with the
  accuracy of about 5~\% but for the Pt dopant it breaks down as the 
  deviations are about 30~\%.

\begin{figure}
\includegraphics[viewport=0.3cm 0.5cm 8.7cm 6.0cm]{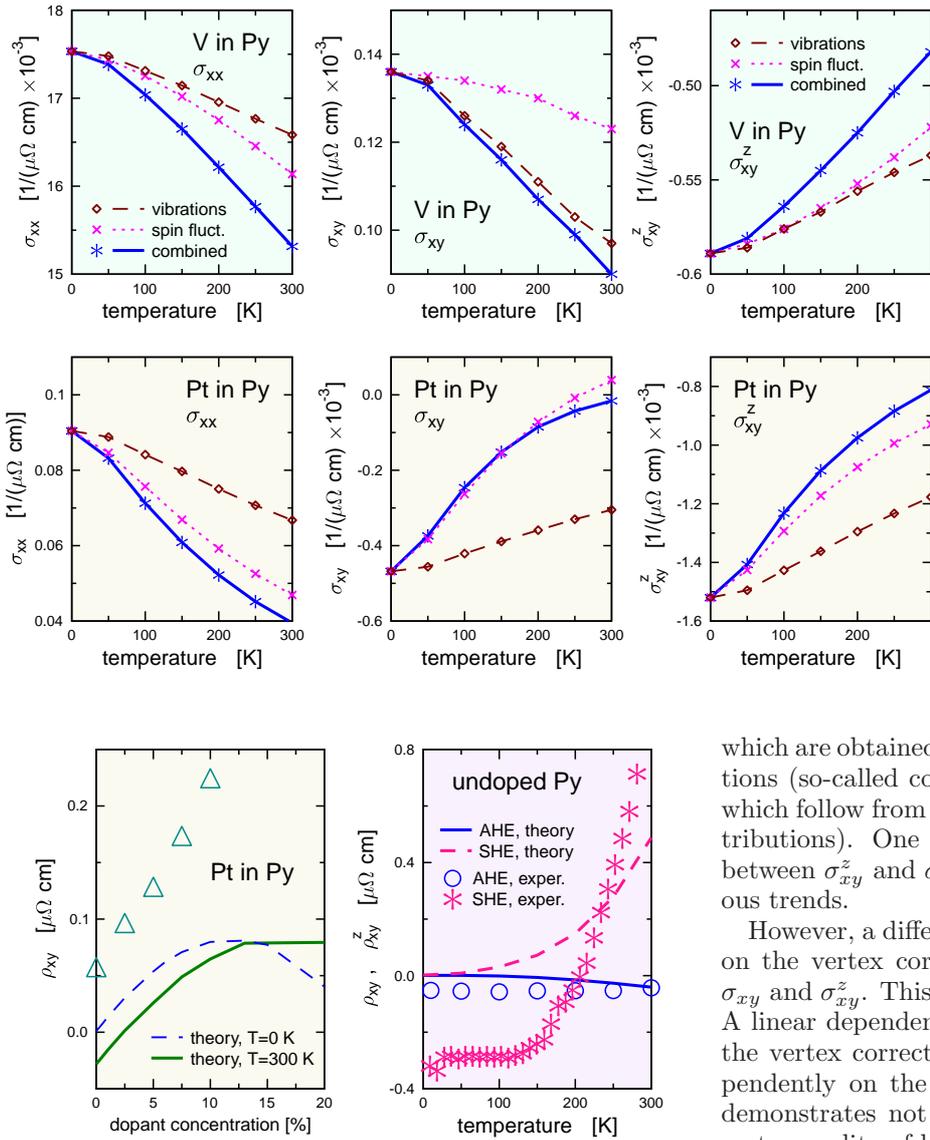}%
\caption{(Color online) Left: Experimental anomalous Hall resistivity
  $\rho_{xy}$ for Pt-doped Py measured by Hrabec
  \ea\ \protect\cite{HGS+16} at 
  $T$=300~K (triangles) together with our theoretical data for $T$=0~K
  (dashed line) and for $T$=300~K (full line).  Right:
  Temperature-dependence of the AHE resistivity $\rho_{xy}$ and the
  SHE resistivity $\rho^{z}_{xy}$ for undoped Py obtained from our
  calculations (lines) and from the experiment of Omori
  \ea\ \protect\cite{OSN+19} (markers).
   Note that the calculations concern bulk Py while
  the experiment \protect\cite{OSN+19} was done for a thin film. }
\label{fig-rho-exp}
\end{figure}

As concerns the comparison of our results with experiment, there are
only few experimental data available for doped Py.  Hrabec
\ea\ \cite{HGS+16} published data on the anomalous Hall resistivity
\rhoxy\ for Pt-doped Py, measured at room temperature.  We compare
their data with our theoretical results for $T$=0~K and for $T$=300~K
in the left panel of Fig.~\ref{fig-rho-exp}.  The agreement is
 worse  
than for the longitudinal transport
(cf.~Fig.~\ref{fig-rho-T300}).  In particular, our calculations
predict a sign-change of \rhoxy\ for room temperature at about 5\%
concentration of Pt, whereas the data of Hrabec \ea\ \cite{HGS+16} do
not exhibit this.  Also the values of \rhoxy\ themselves differ
(albeit they are in the same order of magnitude).  It is not clear
what can be the reason for this.  Let us just note that as concerns
the AMR for Pt-doped Py, the experimental data of Hrabec
\ea\ \cite{HGS+16} differ from the experimental data of Nagura
\ea\ \cite{NST+00} and of Yin \ea\ \cite{YPA+15}.

Calculated temperature-dependence of AHE and SHE resistivities
$\rho_{xy}$ and $\rho^{z}_{xy}$, respectively, for bulk undoped Py
can be compared to experiments done on thin films \cite{OSN+19}.  The
data are shown in the right panel of Fig.~\ref{fig-rho-exp}.  Both
theory and experiment indicate that the SHE resistivity
$\rho^{z}_{xy}$ varies with temperature much strongly than the AHE
resistivity $\rho_{xy}$.  However, despite the general agreement
concerning the trends, there are differences between theory and
experiment concerning particular values, especially for
$\rho^{z}_{xy}$ at low temperatures.  At least part of these
differences certainly is due to the fact that the experiment of Omori
\ea\ \cite{OSN+19} was done for films of 20~nm thickness, meaning that
the experimental data reflect also effects due to the finite thickness
of the film.  In particular, the SHE resistivity $\rho^{z}_{xy}$
measured at $T$=10~K for a 5~nm-thick film was {-2}~\mm{\mu\Omega}cm
whereas for a 20~nm-thick film it was only {-0.3}~\mm{\mu\Omega}cm.
Conjecturely, further increase of the film thickness would decrease
the absolute value of $\rho^{z}_{xy}$ even further, improving thus
the agreement between our theory and experiment.  Note that if the
temperature increases, the agreement between theoretical and
experimental $\rho^{z}_{xy}$ improves.  Presumably, this is because
for high enough temperatures the effects of atomic vibrations and spin
fluctuations will dominate over surface effects.



\subsection{Relation between AHE and SHE}   

\label{sec-ahe-she}

\begin{figure*}
\includegraphics[viewport=0.5cm 0.5cm 18.0cm 6.5cm]{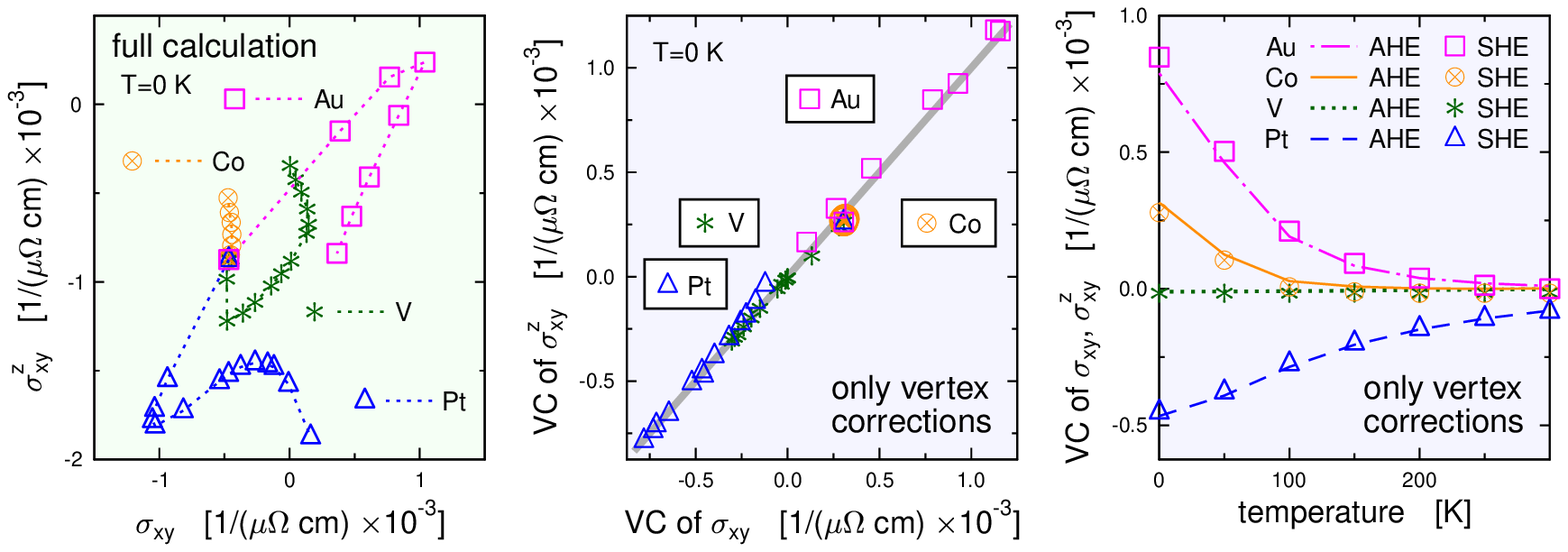}%
\caption{(Color online) Left: Dependence of the SHE conductivity
  \sigZxy\ on the AHE conductivity \sigxy\ for Py doped by V, Co, Pt,
  and Au (of various concentrations) calculated for $T$=0~K if all
  contributions to \sigxy\ and \sigZxy\ are taken into account.
  Middle: As the left panel but considering only the contributions due
  to the vertex corrections.  The straight line corresponds to the
  apparent \mm{\sigma_{xy}\text{(VC)}=\sigma^{z}_{xy}\text{(VC)}}
  relation.  Right: The temperature-dependence of the contributions to
  \sigxy\ (lines) and to \sigZxy\ (markers) due to the vertex
  corrections for Py doped by 6\% of V, Co, Pt, and Au.}
\label{fig-she-ahe}
\end{figure*}

Both AHE and SHE are spin-dependent transport phenomena related to the SOC,
so there is a natural question about their relation.  Following
an initial suggestion of Tsukahara \ea\ \cite{TAK+14}, Omori
\ea\ \cite{OSN+19} argued that within the semiclassical picture and
under some specific assumptions, the skew scattering contributions to
AHE and SHE conductivities are proportional,
\begin{equation}
  \sigma^{\text{skew}}_{xy} \: = \:
  p \,  \sigma^{z,\text{skew}}_{xy}
  \; ,
  \label{eq-polar}
\end{equation}
where $p$ is the spin polarization  of the current,
 \[
  p \: = \:
  \frac{ \sigma_{xx}^{\text{(maj)}} - \sigma_{xx}^{\text{(min)}} }
       { \sigma_{xx}^{\text{(maj)}} + \sigma_{xx}^{\text{(min)}} }
 \; .
 \]  
Exploring the relation between
\sigxy\ and \sigZxy\ for a range of systems like ours might be
instructive.  Therefore, we present in the left panel of
Fig.~\ref{fig-she-ahe} the SHE conductivity \sigZxy\ as a function of
the AHE conductivity \sigxy\ (for $T$=0~K).  We include here all
contributions, i.e., those which are obtained without considering the
vertex corrections (so-called coherent contributions) as well as those
which follow from the vertex corrections (incoherent contributions).
One can see immediately that the relation between \sigZxy\ and
\sigxy\ is quite labyrinthine, with no obvious trends.

However, a different picture emerges if one focuses just on the vertex
corrections (incoherent contributions) to \sigxy\ and \sigZxy.  This
is done in the middle panel of Fig.~\ref{fig-she-ahe}.  A linear
dependence of the vertex corrections to \sigZxy\ on the vertex
corrections to \sigxy\ can be clearly seen, independently on the type
of the dopant.  In fact Fig.~\ref{fig-she-ahe} demonstrates not only
proportionality but an approximate equality of both quantities,
\begin{equation}
    \sigma_{xy}\text{(VC)}  \: = \: \sigma^{z}_{xy}\text{(VC)}
\; .
\label{eq-vc}
\end{equation}
 This relation could be seen as corresponding to
  Eq.~(\ref{eq-polar}), because in our case \mm{p\approx 1}\ (see
  Sec.~\ref{sec-transcal}). 
A complementary view on the relation between the vertex correction
contributions to \sigxy\ and to \sigZxy\ is presented in the right
panel of Fig.~\ref{fig-she-ahe}.  Here we show
\mm{\sigma_{xy}\text{(VC)}}\ and \mm{\sigma^{z}_{xy}\text{(VC)}}\ as
functions of temperature, for Py doped by 6\% of V, Co, Pt, and Au.
Again, Eq.~(\ref{eq-vc}) is satisfied.

Even though
the concept of skew scattering is of limited use
 (and possibly even misleading) 
when dealing
  with changes of the transport properties of Py upon doping 
(see the discussion accompanying
Fig.~\ref{fig-xyonxx}), vertex corrections or, in another terminology,
incoherent contributions \cite{TKD+14} are  robustly
defined even for a
disordered host.  For an ordered host, vertex corrections represent
within our approach the skew scattering \cite{OSN+08}, so we can draw
analogies between the relation Eq.~(\ref{eq-polar}) suggested by Omori
\ea\ \cite{OSN+19} for the skew scattering contributions and
Eq.~(\ref{eq-vc}) satisfied by our data.  Vertex corrections are
related  to scattering-in terms  in the semiclassical Boltzmann
transport theory  \cite{But85}, 
meaning that they reflect extrinsic contributions \cite{OSN+19}.  For
doped Py, their influence on the AHE and SHE is very similar.  What
makes the AHE conductivity \sigxy\ and SHE conductivity
\sigZxy\ different is thus the coherent or intrinsic contribution.



\section{Conclusions}    \label{sec-concl}

Longitudinal charge conductivity \sigxx\ of permalloy decreases with
increasing concentration of the V, Co, Pt, or Au dopants.  This can be
intuitively understood as a consequence of the decrease of
free-electron mean-free path deduced from the broadening of the Bloch
spectral function at \ef.
 The rate of the decrease depends on the dopant type, following
  the sequence Co--Au--Pt--V, in accordance with the scattering
  properties of each atom type. 
Experimental data on the longitudinal
resistivity \rhoxx\ and on the anisotropic magnetoresistance
\mm{(\rho_{zz}-\rho_{xx})/\rho_{\text{aver}}}\ at room temperature
can be 
 qualitatively  
reproduced  in most cases 
  if the effect of finite temperature is included via
the alloy analogy model.
 For the Au dopant, the theoretical and
experimental trends concerning the dependence of the AMR on the
dopant concentration disagree. 

 The calculated dependence of \sigxy\ and \sigZxy\ on the dopant
concentration is  found to be non-monotonic and strongly
depends on the 
temperature.  
The fact that the permalloy host is disordered and not crystalline has
profound influence on how \sigxy\ and \sigZxy\ (characterizing the
anomalous Hall effect and the spin Hall effect, respectively) depend
on the dopant concentration.
In particular, the off-diagonal conductivities \sigxy\ and
\sigZxy\ are not proportional to the longitudinal conductivity
\sigxx\ for low dopant concentrations.  
 As a consequence, the dependence of the AHE and SHE on the
  dopant concentration cannot be ascribed unambigously to skew
  scattering, side-jump scattering, or intrinsic contributions in the
  same way as it can be done when investigating the effect of
  doping for a crystalline host.

The SHE conductivity \sigZxy\ for doped permalloy is not proportional
to the AHE conductivity \sigxy.  However, the vertex corrections to
\sigZxy\ are proportional (and, in fact, approximately equal) to the
vertex corrections to \sigxy.  What makes \sigxy\ and \sigZxy\ of
doped Py different is thus the coherent contribution.

Allowing for the impact of finite temperatures dramatically changes
the overall trends in the dependence of \sigxy\ and \sigZxy\ on dopant
concentrations.  There is no universal rule which of the two effects
we consider, namely, atomic vibrations and spin fluctuations, will be
more important.


\begin{acknowledgments}
This work was supported by the Czech Science Foundation (GA~\v{C}R)
via the project 20-18725S and by the Ministry of Education, Youth and
Sport (Czech Republic) via the project CEDAMNF
CZ.02.1.01/0.0/0.0/15\_003/0000358. Additionally, financial support by
the DFG via SFB~1277 is gratefully acknowledged.
\end{acknowledgments}



\appendix*

\section{Input data for the alloy analogy model}

\label{sec-input}

\begin{table}
\caption{Debye temperatures of elements constituting the systems we
  investigate. Debye temperatures for specific compositions were 
  obtained as weighted averages of the values shown here.}
\label{tab-debye}
\begin{ruledtabular}
\begin{tabular}{lc}
  \multicolumn{1}{c}{element} &
  \multicolumn{1}{c}{$\Theta_{D}$ [K]} \\
\hline
Fe &      420 \\
Ni &      375 \\
V  &      385 \\
Co &      420 \\
Pt &      230 \\
Au &      170 
\end{tabular}
\end{ruledtabular}
\end{table}

\begin{figure}
\includegraphics[viewport=0.5cm 0.5cm 9.0cm 7.5cm]{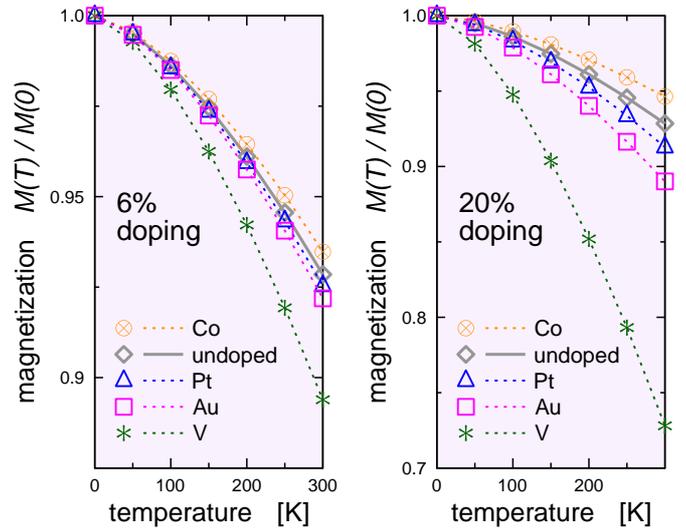}%
\caption{(Color online) Reduced magnetization $M(T)/M(0)$ used as
  input for the alloy analogy model.  Data are shown for undoped Py
  and for Py doped by 6\% or 20\% of V, Co, Pt, and Au.}
\label{fig-M-T}
\end{figure}

We present in this appendix the values we used as input for describing
finite temperature effects by means of the alloy analogy model
(Sec.~\ref{sec-aam}).  The Debye temperatures for each element
constituting the systems we investigate are presented in
Tab.~\ref{tab-debye}.  The reduced magnetization curves
\mm{M(T)/M(0)}\ for two representative concentrations of the dopants
are shown in Fig.~\ref{fig-M-T}.


\bibliography{liter_sigma}

\end{document}